\documentclass{article}
\usepackage{graphicx}
\usepackage{exscale}
\usepackage{amsmath,amsfonts,amssymb}

\newcommand{\mathsize}{}

\DeclareMathSymbol{\symttup}{\mathalpha}{operators}{"0B}
\DeclareMathSymbol{\symttdn}{\mathalpha}{operators}{"0C}
\newcommand{\ttup}{\mathtt\symttup}

\newcommand{\ket}[1]{| #1 \rangle}
\newcommand{\Ket}[1]{\bigl| #1 \bigr\rangle}
\newcommand{\bra}[1]{\langle #1 |}
\newcommand{\Eq}[1]{Eq.~(\ref{#1})}
\newcommand{\Fig}[1]{Fig.~\ref{fig:#1}}
\newcommand{\Sec}[1]{Sec.~\ref{Sec:#1}}
\newcommand{\Uenv}{\mathop{\textstyle{\coprod}}\limits}
\newcommand{\uenv}{\amalg}
\newcommand{\env}{\mathcal E}
\newcommand{\Z}{\mathbb Z}
\newcommand{\C}{\mathbb C}
\newcommand{\F}{\mathbf F}
\newcommand{\sub}{\circleddash}
\newcommand{\Ox}{\mathop{\textstyle{\bigotimes}}\limits}
\newcommand{\op}[1]{\hat{#1}}
\newcommand{\Hil}{\mathcal H}
\newcommand{\Compos}{\mathop{\circledcirc}\limits}
\newcommand{\Prod}{\mathop{\textstyle{\prod}}\limits}
\newcommand{\rev}[1]{\widetilde{#1}}
\newcommand{\Id}{\mathbf 1}
\newcommand{\Swp}{{\mathsf X}}
\newcommand{\Prj}{\op{\mathsf P}}
\newcommand{\subover}[2]{\begin{subarray}{l}#1 \\ #2\end{subarray}}
\newcommand{\dix}[3]{{(\subover{#1}{#2},\,#3)}}
\newcommand{\Q}{{\mathcal Q}}
\newcommand{\Clon}{\complement}
\newcommand{\opClonUp}{\op\Clon\!\ttup}
\newcommand{\Evol}{{\mathbb T}}
\newcommand{\Odot}{\mathop{\textstyle{\bigodot}}\limits}
\newcommand{\Oplus}{\mathop{\textstyle{\bigoplus}}\limits}
\newcommand{\Wedge}{\mathop{\textstyle{\bigwedge}}\limits}

\newcommand{\Alg}{\mathcal A}
\newcommand{\LatPic}[2]{\begin{array}{ccccccc}%
     &\vdots& &\vdots& &\vdots& \\
\cdots& #1 & #2 & #1 & #2 & #1 & \cdots \\
      & #2 &    & #2 &    & #2 \\
\cdots& #1 & #2 & #1 & #2 & #1 & \cdots \\
     &\vdots& &\vdots& &\vdots& \\
\end{array}}

\newcommand{\StI}{\bullet}
\newcommand{\StO}{\circ}
\newcommand{\OO}{\,{}^\StO_\StO}
\newcommand{\OI}{\,{}^\StO_\StI}
\newcommand{\IO}{\,{}^\StI_\StO}
\newcommand{\II}{\,{}^\StI_\StI}
\newcommand{\EMP}{\emptyset}
\newcommand{\UP}{{\uparrow}}
\newcommand{\DN}{{\downarrow}}
\newcommand{\UD}{{\updownarrow}}

\title{On Quantum Cellular Automata}
\date{21 June 2004}
\author{Alexander Yu.\ Vlasov}
\begin{document}
\maketitle
\begin{abstract}
In recent work \cite{SW04} by Schumacher and Werner was discussed
an abstract algebraic approach to a model of reversible quantum cellular 
automata (CA) on a lattice. It was used special model of CA based on 
partitioning scheme and so there is a question about quantum CA derived 
from more general, standard model of classical CA. In present work is 
considered an approach to definition of a scheme with ``history,'' 
valid for quantization both irreversible and reversible classical CA
directly using local transition rules. It is used language of vectors in 
Hilbert spaces instead of $C^*$-algebras, but results may be compared 
in some cases. Finally, the quantum lattice gases, quantum walk 
and ``bots'' are also discussed briefly.
\end{abstract}

\section*{Introduction}

Let us denote Hilbert space of one cell of a quantum cellular (or lattice gas) 
automata as $\Hil_\bullet$, then it is possible to consider different models of 
construction of Hilbert space for whole quantum system. In \cite{SW04} was used 
model with tensor product of different spaces depicted here schematically as
\begin{equation}
\LatPic{\Hil_\bullet}{\otimes},
\label{TensLatH}
\end{equation}
and more abstract model with $C^*$-algebras like
\begin{equation}
\LatPic{\Alg}{\otimes}.
\label{TensLatA}
\end{equation}
It also quite shortly was mentioned other model in relation with quantum lattice
gases and quantum walks \cite[V.E]{SW04}
\begin{equation}
\LatPic{\Hil_\bullet}{\oplus},
\label{PlusLatH}
\end{equation}

The model with tensor products \Eq{TensLatH} is more familiar
in quantum computer science, than model with direct sum \Eq{PlusLatH}. 
In such a case the abstract algebraic approach \Eq{TensLatA} is formally approved, 
because in general case state of cell (or sublattice) of lattice \Eq{TensLatH} may be
simply not defined as a vector in Hilbert space $\Hil_\bullet$, but algebra of 
observables (and state of $C^*$-algebra\footnote{Further, ``state'' always means 
state (ray) in Hilbert space, not state of $C^*$-algebra. The state $\psi\in\Hil$ is 
corresponding to state $\omega_\psi$ of $C^*$-algebra of operators as  
$\omega_\psi \colon \op A \mapsto \bra{\psi}\op A \ket{\psi}$.}) for any
cell(s) is always defined. 

To avoid such a problem here is formally used model \Eq{TensLatH} with finite 
lattices to have correctly defined notion of state for whole system, but quantum gates 
as usually may be defined for arbitrary set of cells (sites). In \Sec{QCA} such a 
system is considered from point of view of usual theory of quantum computational 
networks. The approach is related with question: {\sl how to adopt model of
general CA for quantum computers by using standard tools from theory of quantum 
algorithms}. 

On the other hand, the particular model \Eq{TensLatH} does not necessary 
produce appropriate scheme for description of real space-time processes. 
In \Sec{QLGA} as an illustrative example of physical applications is discussed a 
``qubot'' model of quantum lattice gas automata using both ``additive'' \Eq{PlusLatH} 
and ``multiplicative'' \Eq{TensLatH} schemes. 

\section{Quantum Networks for Cellular Automata}
\label{Sec:QCA}
\subsection{Cellular Automata with ``History''}
\label{Sec:HistCA}

Let us consider usual procedure of rewriting of a classical algorithm for
a quantum computer, {\em i.e.,} two-steps process: 
\begin{enumerate}
\item To change irreversible classical function to reversible one using 
 well known methods: ``quantum function evaluation,''\footnote{Such
 term \cite{QAlg} is used for rather classical method of creation
 reversible function from irreversible one: for function $y = F(x)$
 is considered reversible function on pair of arguments like
 $\rev F\{x,y\} = \{x,F(x) - y\}$, $\rev F\{x,0\}=\{x,F(x)\}$, 
 $\rev F^{-1}=\rev F$, see \Eq{Frn}.\label{fn:qfe}} 
``history tape,'' {\em etc}.
\item To rewrite reversible function, {\em i.e.,} a transposition of a set,
as a matrix of the transposition, {\em i.e.,} the unitary matrix. In such a
way it is possible to write action of the function for arbitrary superposition
of states.
\end{enumerate}

\begin{figure}[htb]
\begin{center}
\includegraphics{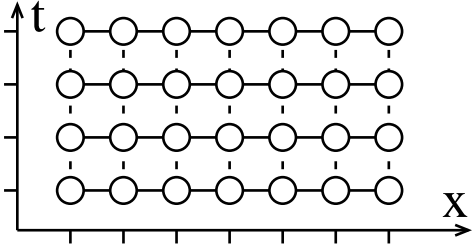}
\end{center}
\caption{Cellular automaton with time (history) dimension} 
\label{fig:ext}
\end{figure}

Here is important to recall, that general definition of classical cellular automata 
includes {\em the time dimension} (see \Fig{ext}) \cite{WCA}
\begin{equation}
 s_i^{(t+1)} = F\bigl(\Uenv_{j \in \env_i} s_j^{(t)}\bigr),
\label{Fn}
\end{equation}
where $s_i \in S$ space of states, $\env_i$ is ``local environment'' of index $i$, 
$\uenv$ is a vector (disjoint union) of states $s_j$ in the environment, and $F$ is
local transition function $F:S^k \to S$. {\em E.g.,} for 
simplest 1D cellular automata used in examples below $F:S^3 \to S$
\begin{equation}
 s_i^{(t+1)} = F(s_{i-1}^{(t)},s_i^{(t)},s_{i+1}^{(t)})
\label{F3}
\end{equation}

Reversible analogue of \Eq{Fn} is map $\rev F : S^{k+1} \to S^{k+1}$, 
$\rev F^{-1} = \rev F$
\begin{equation}
 \rev F \colon \Bigl(s_i^{(t+1)},\Uenv_{j \in \env_i} s_j^{(t)}\Bigr) \mapsto 
 \Bigl( F\bigl(\Uenv_{j \in \env_i} s_j^{(t)}\bigr) \sub s_i^{(t+1)},
 \Uenv_{j \in \env_i} s_j^{(t)}\Bigr)
\label{Frn}
\end{equation}
where $\sub$ is some ``subtracting'' operation $S \times S \to S$,
$a \sub 0 = a$, $a \sub (a \sub b) = b$ like subtraction 
modulo $p$ for $S =\Z_p$ or ``bitwise'' XOR (addition modulo 2) for $S = \Z_2^n$
(see footnote ($^{\ref{fn:qfe}}$) on page \pageref{fn:qfe}). 
For \Eq{F3} map $\rev F : S^4 \to S^4$ may be written as
\begin{equation}
 \bigl(s_i^{(t+1)},s_{i-1}^{(t)},s_i^{(t)},s_{i+1}^{(t)}\bigr) \mapsto
\bigl(F(s_{i-1}^{(t)},s_i^{(t)},s_{i+1}^{(t)}) \sub s_i^{(t+1)},
s_{i-1}^{(t)},s_i^{(t)},s_{i+1}^{(t)}\bigr),
\label{Fr4}
\end{equation}

\begin{figure}[htb]
\begin{center}
\parbox[b]{0.45\textwidth}{
\includegraphics[scale=0.75]{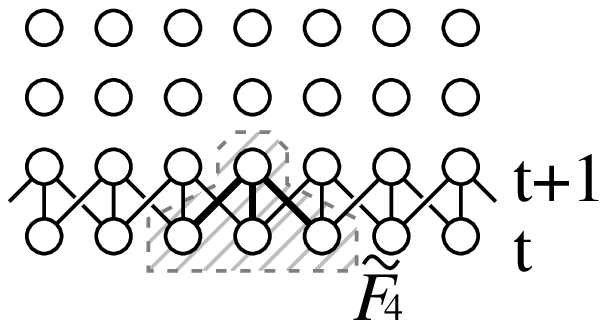}
\bigskip
{\caption{Transition functions}
\label{fig:trans}}
\medskip
}~
\parbox[b]{0.38\textwidth}{
\includegraphics[scale=0.5]{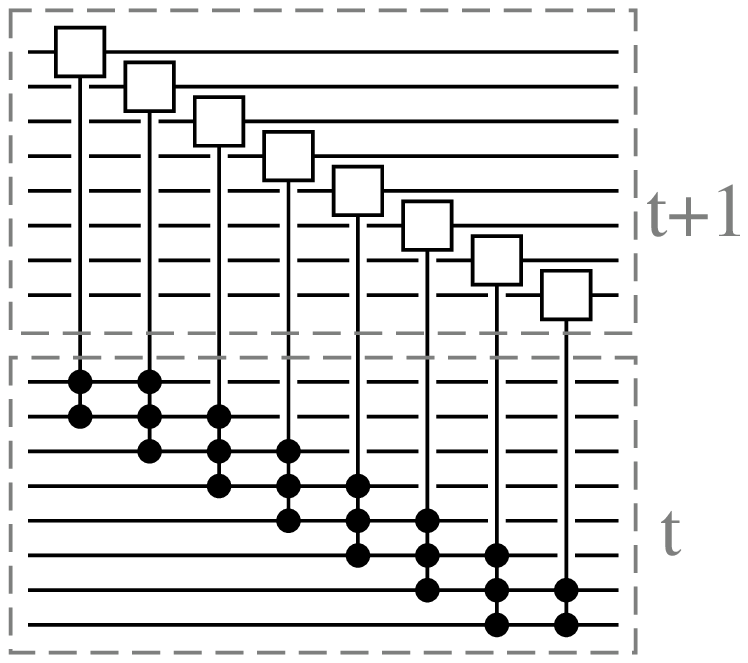}
{\caption{Transition functions represented as network} 
\label{fig:transnet}}
}
\end{center}
\end{figure}

It is clear, that reversible map $\rev F$ at each moment of $t$ acts on two adjacent 
``time layers'' \Fig{trans} and global transition function $\mathbf{\rev F}$ may be 
written as composition of all local $\rev F_i$
\begin{equation}
 \mathbf{\rev F} = \Compos_{i \in L} \rev F_i .
\end{equation}

So instead of one instance of lattice $L$ we have evolution with ``generating
a sequence of copies, history'' as it is quite usual in theory of cellular automata 
\cite{WCA}, reversible classical \cite{Ben73}, and quantum \cite{PB82} computations.

\subsection{Quantum Case}
\label{Sec:HistQCA}

Let us now consider the quantum case. We have some lattice $L$ and
each cell is described by Hilbert space $\Hil_i$,
or $\Hil_i^{(t)}$ with explicit discrete time index
\begin{equation}
\Hil_L = \Ox_{i \in L} \Hil_i,
\quad
\Hil_L^{(t)} = \Ox_{i \in L} \Hil_i^{(t)}.
\end{equation}

Here is suggested that total Hilbert space of evolution may be represented
as tensor product of Hilbert spaces $\Hil_L^{(t)}$ representing lattice
for each time step.
\begin{equation}
 \Hil_\Evol =  \Ox_{t \in \Evol} \Hil^{(t)}_L
\label{tensT}
\end{equation}

The reversible expression \Eq{Frn} may be represented for quantum case
by some unitary operator $\op{F}_i^{(t|t+1)}$ on space $\Hil_{\env_i}^{(t|t+1)}$
\begin{equation}
 \op{F}_i^{(t|t+1)} \colon \Hil_{\env_i}^{(t|t+1)} \to \Hil_{\env_i}^{(t|t+1)};
 \quad  \Hil_{\env_i}^{(t|t+1)}=
  \Hil_i^{(t+1)}\otimes\bigl(\Ox_{j \in \env_i} \Hil_j^{(t)}\bigr),
\label{HilLocF}
\end{equation}
expressed more directly as
\begin{equation}
\op{F}_i^{(t|t+1)} = ~\sum_{\rlap{$\scriptstyle s_1,\dots,s_k \in S$}}~ 
 \op f^{(i,t+1)}_{s_1,\dots,s_k}%
 \otimes \Prj_{s_1}^{(i_1,t)}\otimes \cdots \otimes \Prj_{s_k}^{(i_k,t)},
 \quad \{i_1,\dots,i_k\} \in \env_i,
\label{ExprLocF}
\end{equation}
where $\Prj_{s}$ are projectors, $\op f$ are operators corresponding to local 
transition table representing $F$ of the cellular automata (say one of two 
$2 \times 2$ matrices $\op\Id$ and $\op\sigma_x$ in simplest case of $S = \Z_2$), and
upper index $\op\cdot^{(i,t)}$ of a ``one-site'' operator like $\Prj$ or $\op f$ 
corresponds to space-time coordinate $(x,t)$, see \Fig{ext}.

The global transition function $\op{\F}^{(t \mapsto t+1)}$ 
may be represented as
\begin{equation}
 \op{\F}^{(t \mapsto t+1)} = \Prod_{i \in L} \op{F}_i^{(t|t+1)}.
\label{locprod}
\end{equation}
The expression \Eq{locprod} is valid, because $\op{F_i}$ commute despite of overlapping
domains (see \Fig{trans}). It is more clear from representation of the global
function $\op{\F}$ as a quantum network \Fig{transnet} where $\op{F}_i$ 
are ``{\sf Controlled-F}'' gates and the commutativity may be checked by 
straightforward calculations using \Eq{ExprLocF}.

\subsection{``Histories'' Entanglement}
\label{Sec:HistEnt}

The specific property of given model is {\em entanglement between states
at different times}, see \Eq{FtEnt} below. 
It is related with some difference of considered
scheme and alternative approach. Let us consider {\em global}
transition rule
\begin{equation}
 \op{\F}^{(t \mapsto t+1)} \colon 
 \Hil_L^{(t)} \otimes \Hil_L^{(t+1)} \to \Hil_L^{(t)} \otimes \Hil_L^{(t+1)}.
\label{globrul}
\end{equation}
It has certain difference with ``na\'\i ve evolution'' approach
\begin{equation}
 \op{\F}_e^{(t \mapsto t+1)} \colon 
 \Hil_L^{(t)} \to \Hil_L^{(t+1)}
\label{globold}
\end{equation}
or even 
\begin{equation}
 \op{\F}_e \colon 
 \Hil_L \to \Hil_L.
\label{globone}
\end{equation}

In the \cite{SW04} is used Heisenberg picture instead of Schr\"odinger one
used here, but all expressions used above may be simply rewritten in such
a picture
\begin{equation}
 \op A \mapsto  \op{\F} \op A \op{\F}^{-1}
\end{equation}
and it is clear, that in \cite{SW04} is used approach with \Eq{globone}, but not
with \Eq{globrul}. 

On the other hand, \Eq{globold} or \Eq{globone} may be rewritten
in form \Eq{globrul} using special operator defined on basis elements as
\begin{equation}
 \op{\F}_\Swp\bigl(\ket{K} \otimes \ket{R}\bigr) = 
 \ket{R} \otimes (\op{\F}_e\ket{K}),
\quad
\op{\F}_\Swp = \op{\Swp}_{12} \circ (\Id \otimes \op{\F}_e),
\label{RLEswap}
\end{equation}
where $\op{\Swp}_{12}$ is ``swap'' operator. 

It is suggested, that second space is ``initialized'' by some state $\ket{0}$,
and it is clear that already for classical global transition function $\F$
we have two different schemes. The scheme discussed in present paper is defined 
on basic states as
\begin{equation}
\op{\F}\colon \ket{K} \otimes \ket{R} \mapsto 
\ket{K}\otimes \ket{\F(K) \sub R},
\quad
\op{\F}\Ket{K,0} = \Ket{K,\F(K)}
\label{Fproc}
\end{equation}
and unitary both for reversible and
irreversible global functions. Let us consider application of \Eq{Fproc}
to {\em composition} of two basic states of lattice
\begin{equation}
\op{\F}\bigl((\alpha\ket{K_1}+\beta\ket{K_2}) \otimes \ket{0}\bigr) 
= \alpha\ket{K_1}\otimes\ket{\F(K_1)}
 +\beta\ket{K_2}\otimes \ket{\F(K_2)}
\label{FtEnt}
\end{equation}
The \Eq{FtEnt} describes entangled state if  $\F(K_1) \ne \F(K_2)$,
{\em i.e.,} for reversible CA such states are always entangled and only for 
irreversible CA some compositions are evolving to non-entangled states.

Other scheme, \Eq{RLEswap}
\begin{equation}
\op{\F}_\Swp\Ket{K,0} = \Ket{0,\F_e(K)}
 \tag{\ref{RLEswap}$'$}
\label{Fevol}
\end{equation}
unitary only for reversible global function.  It is clear, that \Eq{Fevol} never
entangles two states
\begin{equation}
\op{\F}_\Swp\bigl((\alpha\ket{K_1}+\beta\ket{K_2}) \otimes \ket{0}\bigr) 
= \ket{0} \otimes \bigl(\alpha\ket{\F(K_1)}+\beta\ket{\F(K_2)}\bigr)
\label{FxNoEnt}
\end{equation}
and so may be really modeled by simpler expression
\Eq{globone}. On the other hand \Eq{Fproc} entangles two terms, but
for reversible function it may be disentangled using function%
\footnote{It should be mentioned, that $\op{\F}'$ disentangles result
of application of $\op{\F}$ if second state is $\ket{0}$. In more
general case {\mathsize$ \op{\F}' \bigl(\op{\F}(\ket{K}\ket{R})\bigr) 
 = \ket{\F^{-1}(\F(K)\sub R)\sub R}\ket{\F(K)\sub R}$}.}
\begin{equation}
\op{\F}'\colon \ket{K} \otimes \ket{R} \mapsto 
\ket{\F^{-1}(R)\sub K}\otimes \ket{R},
\label{Finv}
\end{equation}
\begin{equation}
\op{\F}'\Ket{K,\F(K)}= \Ket{0,\F(K)}
\quad \Rightarrow \quad
 \op{\F}' \bigl(\op{\F}(\ket{\psi}\ket{0})\bigr)=\ket{0}\bigl(\op\F_e\ket{\psi}\bigr).
\label{Ferase}
\end{equation}

\subsection{Second-Order CA}
\label{Sec:IICA}

\begin{figure}[htb]
\begin{center}
\includegraphics{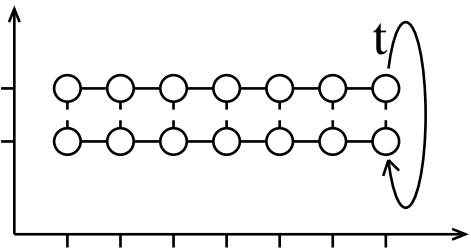}
\end{center}
\caption{Cyclic time condition (with two layers)} 
\label{fig:cyctime}
\end{figure}

Yet another interesting possibility --- is to consider suggested model of
cellular automata with cyclic time. For such a case it is also possible
to rewrite \Eq{globrul} as \Eq{globone} for some reversible CA.
Let us consider for example simple case with two time steps \Fig{cyctime} 
(here second term in \Eq{Fproc} already is not always $\ket{0}$).

In such a case evolution \Eq{globrul} may be expressed in form \Eq{globone} using 
new model with same lattice, but extended configuration space
of each cell $\Hil''_i = \Hil_i \otimes \Hil_i$. In classical case
it corresponds to well known ``second-order'' Fredkin scheme for construction 
of reversible cellular automaton from irreversible one using two consequent states 
of lattice for calculation of each step, similarly with some reversible second-order
differential equations \cite{WCA,TM90}.

\begin{figure}[htbp]
\begin{center}
\includegraphics[scale=0.75]{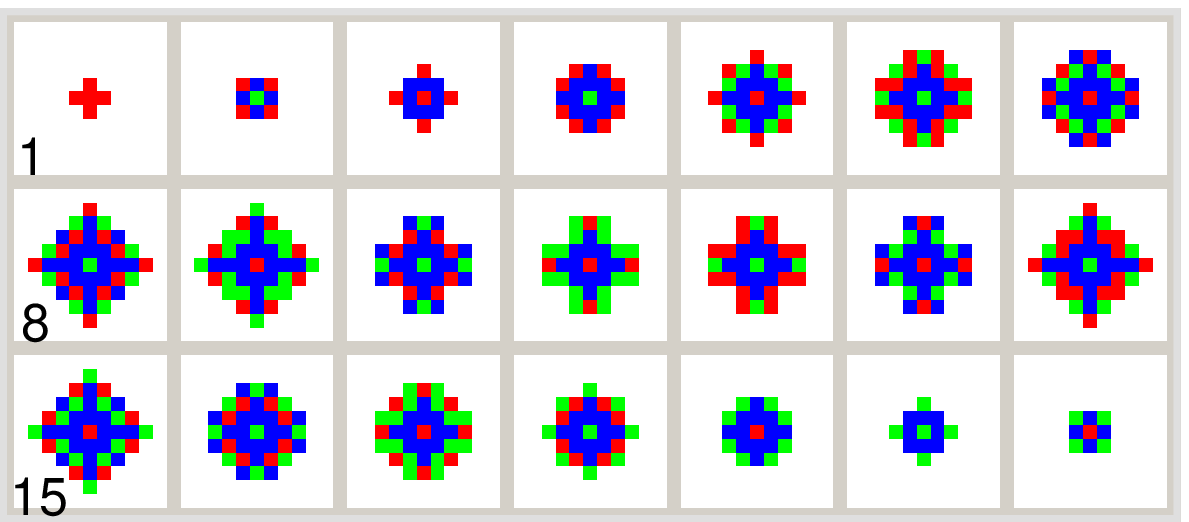}

\includegraphics[scale=0.75]{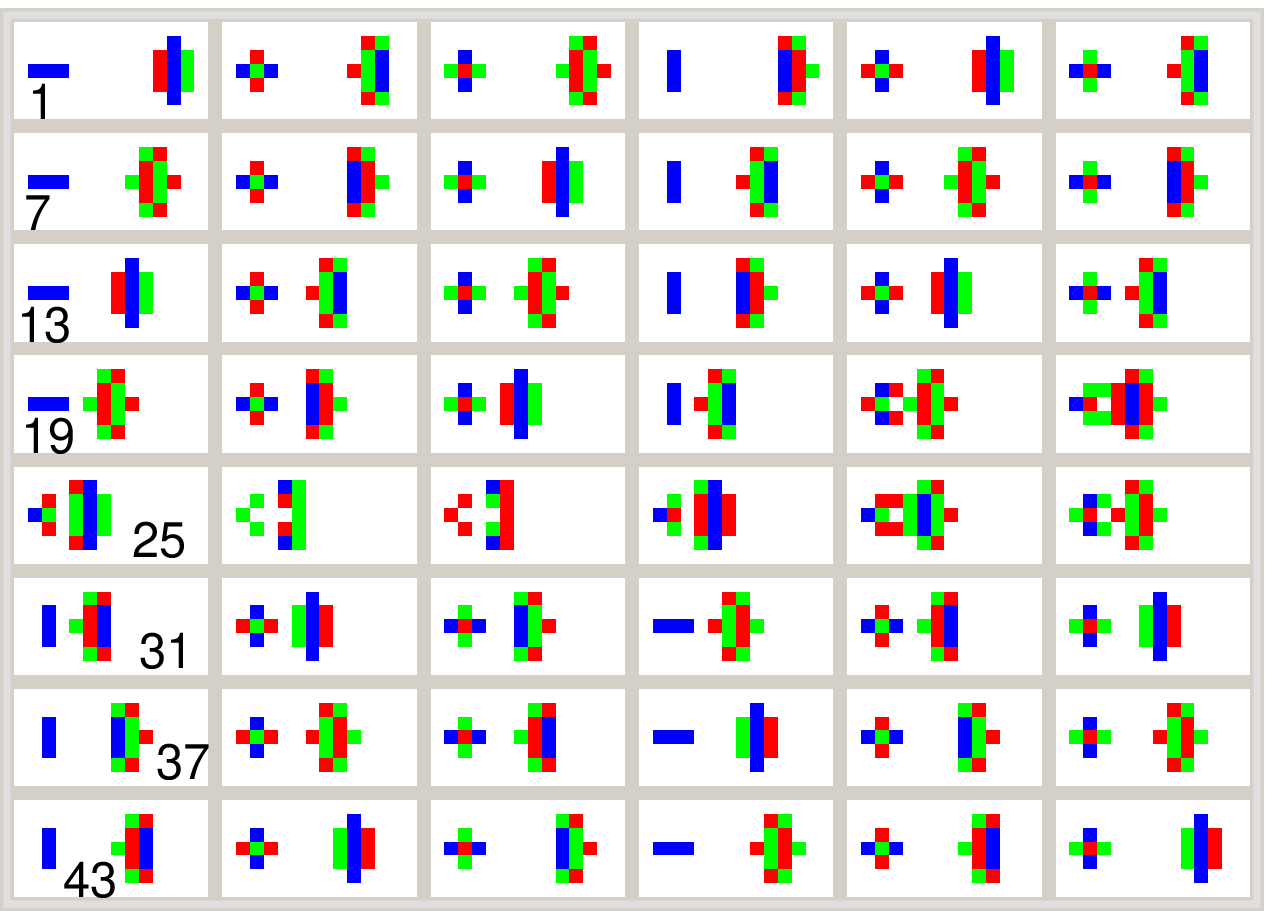}

\includegraphics[scale=0.75]{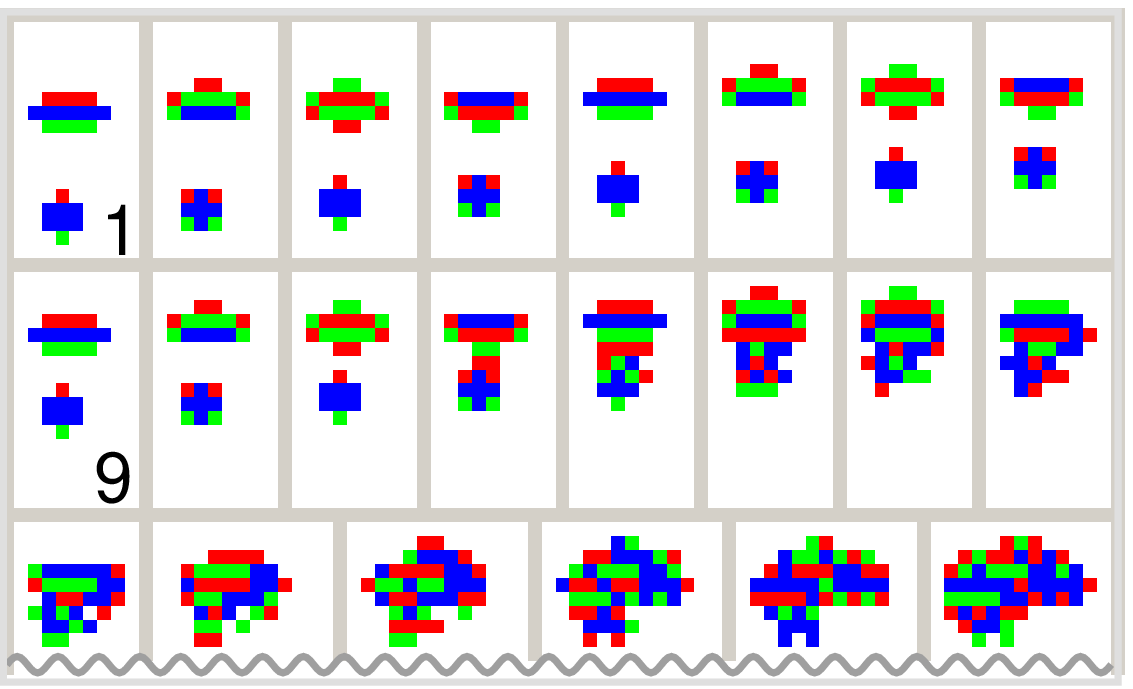}
\end{center}
\caption{``Reversible Game of Life'' examples} 
\label{fig:rep}
\end{figure}

On \Fig{rep} is reproduced example of evolution of reversible classical CA
produced by such a way from famous Conway's ``Game of Life'' irreversible 
CA \cite{Life}. 
Methods described above let us consider action of such CA
on arbitrary quantum superposition of basic states, and generalization
to ``quantum transition tables'' is more or less straightforward, but
outside of scope of this note.
 
\begin{figure}[htb]
\begin{center}
\includegraphics{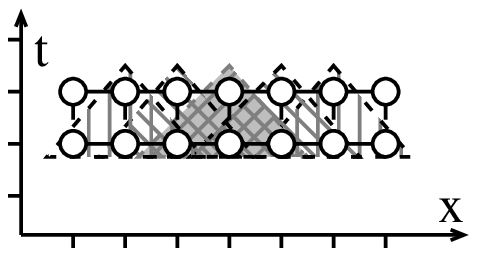}
\includegraphics{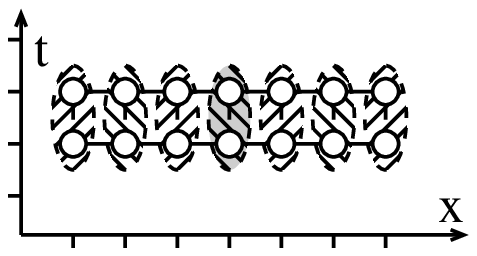}
\end{center}
\caption{Formal partitioning for \Fig{cyctime}} 
\label{fig:cycpart}
\end{figure}

It is also possible to use the same internal space, but double lattice $L \times L$,  
$\Hil_L \otimes \Hil_L \cong \Hil_{L \times L}$. 
Evolution is described by two steps, with second one is swap of two copies
of lattice\footnote{Or ``parts'' of internal state in approach with extended
space.}, but unlike \cite{SW04} formal partitioning scheme used for such process 
\Fig{cycpart} has overlapped partitions at first step ($F_i$, see also \Fig{trans}).

\subsection{Two-Steps Partitioning}
\label{Sec:PartCA}

\begin{figure}[htb]
\begin{center}
\parbox[t]{0.45\textwidth}{
\includegraphics[scale=0.75]{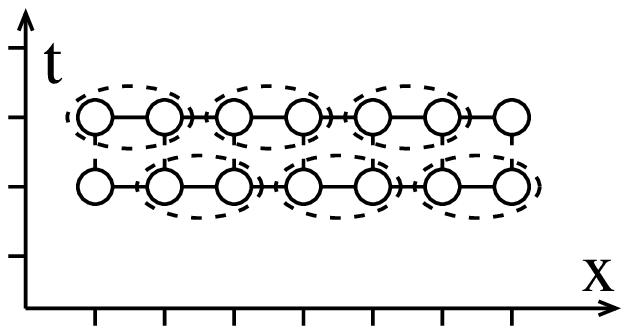}
{\caption{Two-step partitioning}
\label{fig:margpart}}
}~
\parbox[t]{0.45\textwidth}{
\includegraphics[scale=0.75]{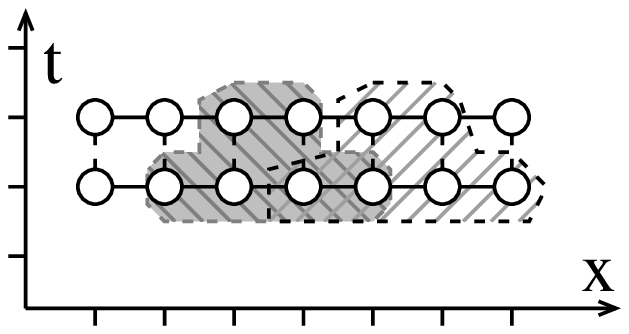}
{\caption{Extended partitioning} 
\label{fig:extpart}}
}
\end{center}
\end{figure}

Scheme suggested above still does not include Margolus partitioning used in 
\cite{SW04} and described as two-step process \Fig{margpart}. More rigorously,
Margolus scheme as particular example of
definition \Eq{Fn} (with two different $F$ for odd and even moments of time)
is valid for quantization of classical CA with transition function $F$, but does 
not have extension to quite straightforward generalization to quantum case with 
general unitary operator $\op \Q$ applied to each subunit of partitioning 
(dashed ellipse in example with 1D cellular automaton on \Fig{margpart}).

The model needs for subtler construction. It is possible for example to
consider extended $(t,t+1)$ neigbourhood (see \Fig{extpart}) for local transition 
functions $\op{F}_{i \times n}^{(t|t+1)}$  and use more difficult expression
instead of \Eq{ExprLocF}.

Let us consider example with 1D cellular automata with Margolus partitioning 
and unitary transformation  $\op\Q : \Hil \otimes \Hil \to \Hil \otimes \Hil$ 
applied two times to different partitioning \Fig{margpart}.
Then local transition function $\op{F}_{i \times 2}^{(t|t+1)}$ for 
{\em both steps of the process} using neighborhood with 6 ($4 \to 2$) elements 
in two time layers is depicted on \Fig{extpart} and may be expressed as
\begin{equation}
\begin{split}
&\op{F}_{i\times 2}^{(t|t+1)} 
 = \bigl(\op \Q^\dix{2i-1}{2i}{t} \otimes \op \Q^\dix{2i+1}{2i+2}{t}\bigr)^{\!-1}
\op \Q^\dix{2i}{2i+1}{t+1} \opClonUp 
\bigl(\op \Q^\dix{2i-1}{2i}{t} \otimes \op \Q^\dix{2i+1}{2i+2}{t}\bigr), \\
&\opClonUp 
=\op{\Clon}^\dix{2i}{2i+1}{\subover{t}{t+1}}
= \sum_{s_1,s_2}
 (\op{\mathsf U}^{(2i,t+1)})^{s_1}\otimes (\op{\mathsf U}^{(2i+1,t+1)})^{s_2}\otimes
 \Prj^{(2i,t)}_{s_1} \otimes \Prj^{(2i+1,t)}_{s_2},
\label{SuperF}
\end{split}
\end{equation}
where $\opClonUp$ acts on basis elements as
\begin{equation}
 \opClonUp \colon
 \Ket{s_1,s_2}\otimes \Ket{0,0} \mapsto \Ket{s_1,s_2}\otimes \Ket{s_1,s_2},
\label{Clon2}
\end{equation}
and $\op{\mathsf U}$ in \Eq{SuperF} is Weyl ``cyclic shift'' operator,
$\op{\mathsf U}: \ket{s} \mapsto \ket{s + 1 \mod m}$.

So at first $\op \Q$ are applied to blocks $(2i-1,2i|t)$ and $(2i+1,2i+2|t)$,
next, $\opClonUp$ ``spreads'' block $(2i,2i+1)$ from $(t)$ to $(t+1)$, 
$\op \Q$ is applied to block $(2i,2i+1|t+1)$ and, finally, first two 
applications of $\op \Q$ to $(2i-1,\cdots,2i+2|t)$ are ``undone''. 
The formula \Eq{SuperF} let us check directly, that such operators 
$\op{F}_{2\times i}^{(t|t+1)}$ 
for different ``even neighborhoods'' \Fig{extpart} are really commuting.

It should be mentioned, that such definition of transition function depends 
on choice of basis in Hilbert space, because definition of $\opClonUp$
\Eq{Clon2} depends on basis --- in agreement with famous no-cloning theorem 
\cite{noclon} only set of orthogonal states may be cloned perfectly.
On the other hand, it may be simply checked that for change of basis in each
cell described by unitary operator $\op{\mathcal B}$ we formally simply
may use the same definition of $\op{F}_{i\times 2}^{(t|t+1)}$ with
new function
\begin{equation}
\op\Q' = (\op{\mathcal B} \otimes \op{\mathcal B}) \: \op\Q \:
(\op{\mathcal B}^* \otimes \op{\mathcal B}^*).
\end{equation}

Furthermore, it is possible to consider new lattice with two-cells block of 
partitioning (at second time step) considered as new cell of lattice. Then we have
new transition function $\op{F}_i$ with ``shape'' \Fig{trans} like \Eq{HilLocF},
but expression more difficult than \Eq{ExprLocF}.
Such observation let us suggest for definition of quantum cellular automata
arbitrary set of local transition functions with only condition
\begin{equation}
 \op{F}_i^{(t|t+1)} \op{F}_j^{(t|t+1)} = \op{F}_j^{(t|t+1)} \op{F}_i^{(t|t+1)},
\quad
[\op{F}_i^{(t|t+1)},\op{F}_j^{(t|t+1)}]=0,
\end{equation}
or maybe even more general
\begin{equation}
 \op{F}_i^{(t|t+1)} \op{F}_j^{(t|t+1)} = 
\omega_{i,j}\op{F}_j^{(t|t+1)} \op{F}_i^{(t|t+1)},
\quad |\omega_{i,j}|=1,
\end{equation}
because common complex phase does not change state and so global transition 
function represented as product \Eq{locprod} is correct for any ordering
of $\op{F}_i$.

\subsection{Problem with Space-Time QCA Models}
\label{Sec:ProbCA}

The model of QCA considered here may be realized using standard quantum
network model, but in such a case the history is implemented not as time
dimension, but as additional dimension of ``hypercube network'' necessary
for ``quantum function evaluation.'' Is it possible to use such
QCA as models of some physical processes in space-time?

Spacetime localised algebras was briefly discussed in 
\cite[V.F]{SW04}. 
With approach used in present paper construction of QCA used in
\cite{SW04} may be compared with a model of reversible CA 
``erasing their own history'' via $\op{\F'}$ \Eq{Ferase}. If
it is really necessary to perform such erasure, especially if to keep in
mind possibility of applications to theory to irreversible CA?
Moreover, in initial expression \Eq{Fn} for local classical transition rule there 
is no clear difference between reversible and irreversible case.

There is certain problem with {\em covariance} for QCA with space-time 
lattice as a model of physical events. Let us consider for example ``lattice'' 
with only one point and two states. How to model even trivial evolution with 
spreading without change?
It is not possible to use map like
\begin{equation}
\ket{\psi}\underbrace{\ket{0}\cdots\ket{0}\!}_T \to 
\ket{\psi}\ket{\psi}\underbrace{\ket{0}\cdots\ket{0}\!}_{T-1} \to 
\ket{\psi}\ket{\psi}\ket{\psi}\underbrace{\ket{0}\cdots\ket{0}\!}_{T-2} \to  \cdots ,
\label{Tclon}
\end{equation}
unless $\ket{\psi}$ is not one of two fixed orthogonal states, as it was
always in consideration above, because otherwise \Eq{Tclon} describes 
{\em nonlinear} map, it is the subject of quantum {\em no-cloning} 
theorem \cite{noclon}. 

Minor problem here is non-invariant state $\ket{0}^{\otimes T}$, because 
formally it may be corrected by addition of third, ``empty (vacuum) state''
$\ket{\emptyset}$, but it does not resolve main problem, because
\begin{equation}
\ket{\psi}\ket{\emptyset}^{\otimes T} \to 
\ket{\psi}\ket{\psi}\ket{\emptyset}^{\otimes T-1} \to 
\ket{\psi}\ket{\psi}\ket{\psi}\ket{\emptyset}^{\otimes T-2}  \cdots 
\tag{\ref{Tclon}$'$}
\end{equation}
is nonlinear cloning anyway and so prohibited by quantum laws.

The problem is not only due to suggested approach, it was already mentioned,
that ``na\'\i{}ve evolution'' model may be described in similar way.
In such a case instead of \Eq{Tclon} we would write 
\begin{equation}
\ket{\psi}\ket{0}^{\otimes T} \to \ket{0}\ket{\psi}\ket{0}^{\otimes T-1} 
 \to \ket{0}\ket{0}\ket{\psi}\ket{0}^{\otimes T-2}  \cdots.
\label{Teras}
\end{equation}
The \Eq{Teras} with ``automata erasing own histories'' is certainly linear,
unitary, but it is not a picture we could expect for description of space-time
model of real physical system.

Yet another idea is to use direct sum instead of tensor product for ``joining''
of state of lattice for different times
\begin{equation}
 \Hil^{\oplus}_\Evol =  \Oplus_{t \in \Evol} \Hil^{(t)}_L,
\label{plusT}
\end{equation}
but it is not clear from very beginning, why we should distinguish time dimension by
such a way, especially in applications for relativistic models.
  
Some clarification of the question may be based on application of 
quantum lattice gas automata (QLGA) model and discussed in \Sec{qlgt}.
Simplest model of transition from QLGA to QCA is considered in 
\Sec{QLG2CA}. This example prompts yet another possible representation
for one-particle trivial evolution as composition (up to normalization)
\begin{equation}
\ket{\psi^\Evol} =
\ket{\psi}\ket{0}^{\otimes T} + \ket{0}\ket{\psi}\ket{0}^{\otimes T-1} 
 + \ket{0}\ket{0}\ket{\psi}\ket{0}^{\otimes T-2}  + \cdots,
\label{Tcomp}
\end{equation}
but already for two particles trivial evolution, an expression may
be rather cumbersome \Eq{TrivTwo}.

\section{Quantum Lattice Gas Automata (QLGA)}
\label{Sec:QLGA}

From point of view of physical applications the theory of {\em lattice gases} 
\cite{WCA,TM90,Mey96,MeyQLGA} devotes special attention.
For short ``translation'' of some ideas of quantum field theory to language of 
quantum information science, here may be convenient to use following model. 

\subsection{Quantum `Bot' on Lattice}

The {\em quantum bot} or {\em qubot} \cite{qubot} is quantum system with 
Hilbert space decomposed in natural way on two components:
\begin{equation}
\Hil_\bot=\Hil_\ell \otimes \Hil_S,
\end{equation}
there $\Hil_\ell$ corresponds to spatial
degrees of freedom of lattice ($\dim \Hil_\ell =l=k^D$ for D-dimensional 
hypercubic lattice with $k$ cells in each side) and $\Hil_S$ --- to internal states. 
It is {\em the programmed quantum excitation}, just approach with model \Eq{PlusLatH}, 
and also has analogue with {\em quantum robots} \cite{BQR1}.

\begin{figure}[htb]
\begin{center}
\includegraphics[scale=0.75]{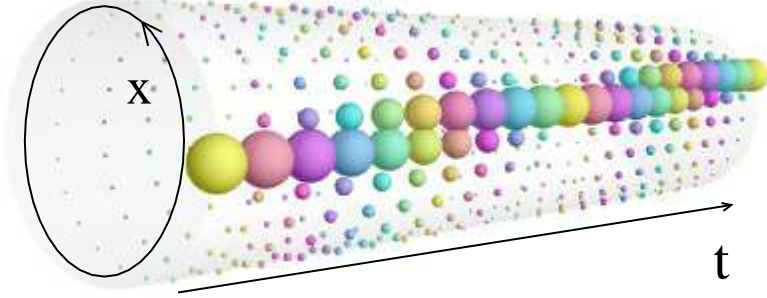}
\end{center}
\caption{Evolution of a qubot on cycle (size -- amplitude, color -- phase)} 
\label{fig:qubot}
\end{figure}

Evolution of qubot may be described by {\em conditional quantum dynamics} 
\cite{BDEJ95}, a simple case with $\dim \Hil_S = 2$ is 
\begin{equation}
 \op E_\bot = \mathsf U \otimes \ket{0}\bra{0} + \mathsf U^* \otimes \ket{1}\bra{1},
\label{RLbot}
\end{equation}
where $\mathsf U$ is Weyl shift operator, {\em i.e.,} for internal state 
$\ket{0}$ or $\ket{1} \in \Hil_S$ \Eq{RLbot} describes either left or right 
translation on the lattice $\Hil_\ell$. 
For simple expression \Eq{RLbot} it is even possible to find Hamiltonian or
consider ``continuous time evolution'' $\op E_\bot^\tau$ \cite{qubot}, see \Fig{qubot}.

So-called {\em coined quantum walk (CQW) on cycle} \cite{qwolf} may be described
as composition of \Eq{RLbot} and Hadamard transform applied to $\Hil_S$. 
Really it is not quite clear, if Hadamard CQW may be considered as ``true quantum 
analogue'' of classical random walk --- it rather resembles superposition of two 
excitations (``qubots'') traveling in opposite directions. It is especially
clear, if to choose new basis in $\Hil_S$, there Hadamard transform becomes diagonal.
Such a note may be essential, {\em e.g.,} 
quantum walk with proper correspondence with classical case has straightforward
representation using infinite-dimensional internal space $\Hil_S$, but it
should be discussed elsewhere. Furthermore, CQW is not only suggested
model of quantum walk \cite{expwalk}, and most likely it was discussed in 
\cite[V.E]{SW04} just due to the natural tie with lattice gas models.

\subsection{Systems with Many Qubots}

The total Hilbert space with $n$ equal qubots on a lattice may be described
as symmetric product
\begin{equation}
 \Hil_{\top^n}=\Hil_{\odot^n} = \Odot_{i=1}^n \Hil_\bot
\label{squbot}
\end{equation}
or as antisymmetric one
\begin{equation}
 \Hil_{\top^n} = \Hil_{\wedge^n} = \Wedge_{i=1}^n \Hil_\bot.
\label{aqubot}
\end{equation}

\begin{figure}[htb]
\begin{center}
\parbox[t]{0.3\textwidth}{
\includegraphics[scale=0.75]{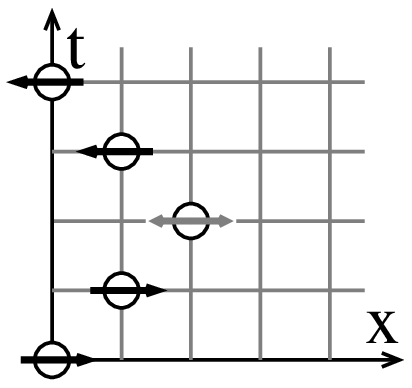}\\
 a) $\Hil_{\bot\mathnormal 1}$
}~
\parbox[t]{0.3\textwidth}{
\includegraphics[scale=0.75]{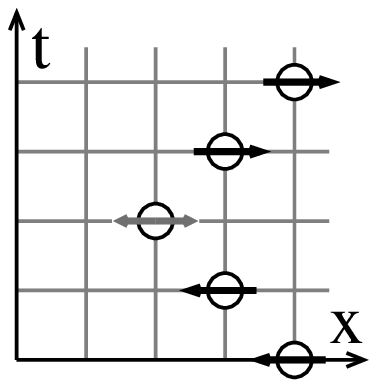}\\
 b) $\Hil_{\bot\mathnormal 2}$
}~
\parbox[t]{0.3\textwidth}{
\includegraphics[scale=0.75]{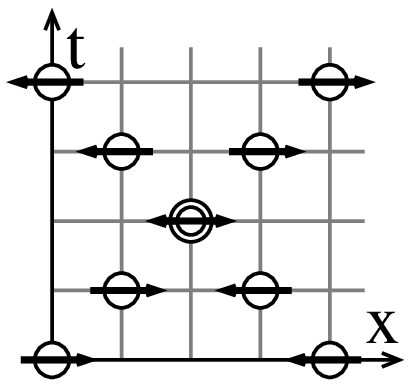}\\
 c) $\Hil_{\bot\mathnormal 1} \odot \Hil_{\bot\mathnormal 2}$
}
{\caption{Reversible QLGA (symmetric product)} 
\label{fig:lga}}
\end{center}
\end{figure}

\begin{figure}[htb]
\begin{center}
\parbox[t]{0.3\textwidth}{
\includegraphics[scale=0.75]{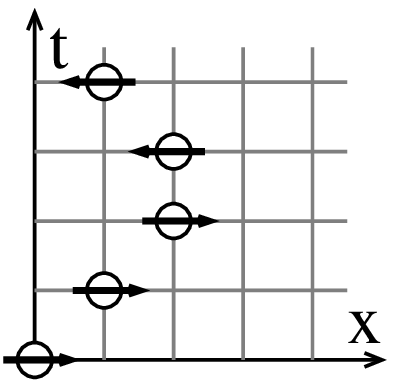}\\
 a) $\Hil_{\bot\mathnormal 1}$
}~
\parbox[t]{0.3\textwidth}{
\includegraphics[scale=0.75]{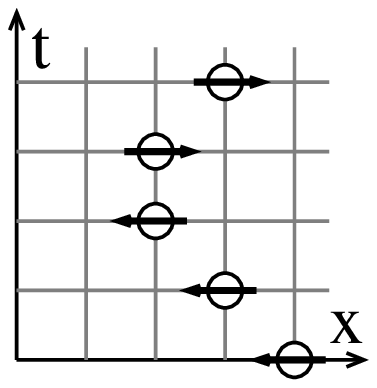}\\
 b) $\Hil_{\bot\mathnormal 2}$
}~
\parbox[t]{0.3\textwidth}{
\includegraphics[scale=0.75]{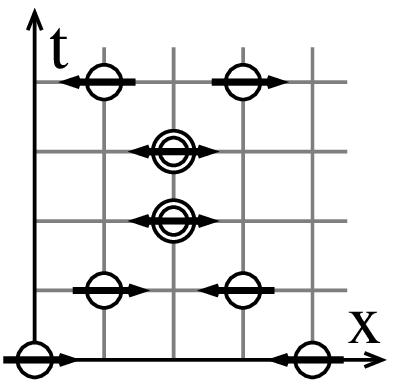}\\
 c) $\Hil_{\bot\mathnormal 1} \otimes \Hil_{\bot\mathnormal 2}$
}
{\caption{Reversible QLGA (tensor product)} 
\label{fig:lga0}}
\end{center}
\end{figure}

Simpler expression with usual tensor product
\begin{equation}
 \Hil_{\top^n} = \Hil_{\otimes^n} = \Ox_{i=1}^n \Hil_\bot,
\label{dqubot}
\end{equation}
does not take into account quantum statistics and may be used for description
of $n$ distinguishable qubots or as preliminary step for construction of more
complicated expressions \Eq{squbot} and \Eq{aqubot}. It should me mentioned
yet, that often quantum indistingushability principle may be essential.
See for example \Fig{lga}, where at moment of collision of particles 
state of composite system is defined correctly \Fig{lga}(c), but state of each 
particular particle formally may be undefined \Fig{lga}(a,b). It is even
does not clear, if \Fig{lga} describes elastic collision or noninteracting
particles.

\begin{figure}[htb]
\begin{center}
\parbox[t]{0.3\textwidth}{
\includegraphics[scale=0.75]{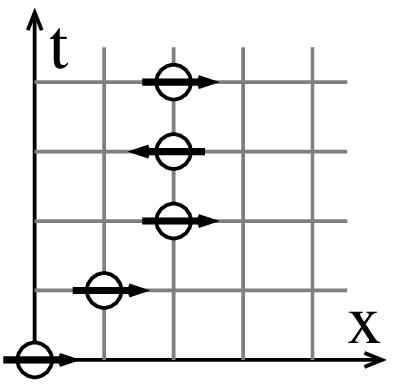}\\
 a) $\Hil_{\bot\mathnormal 1}$
}~
\parbox[t]{0.3\textwidth}{
\includegraphics[scale=0.75]{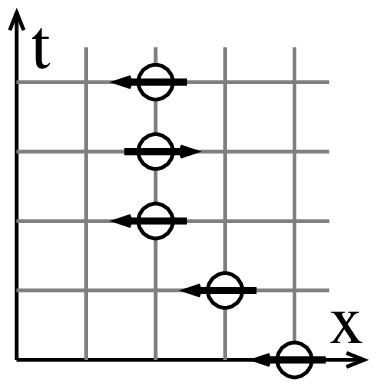}\\
 b) $\Hil_{\bot\mathnormal 2}$
}~
\parbox[t]{0.3\textwidth}{
\includegraphics[scale=0.75]{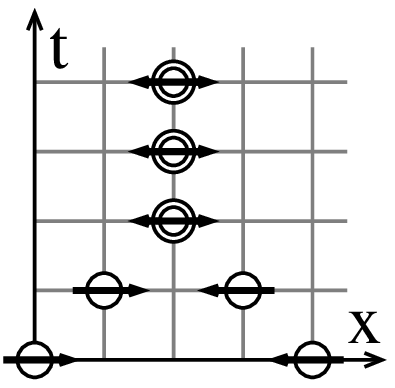}\\
 c) $\Hil_{\bot\mathnormal 1} \odot \Hil_{\bot\mathnormal 2}$
}
{\caption{``Pseudo-irreversible'' process with QLGA} 
\label{fig:ilga}}
\end{center}
\end{figure}

On the other hand, ``a phase shift'' due to interaction on \Fig{lga0} may be modelled 
using tensor product of two ``slightly'' nonequivalent qubots, but it is also
possible with symmetric product of two qubots with extended internal space
for counting time of ``clinch''. 

With infinite-dimensional internal space it is possible to make the time
of ``clinch'' infinite, {\em i.e.,} model process like non-elastic collision, 
usually considered as irreversible \Fig{ilga}.

\medskip

{\em The Fock space} for system with varying number of qubots may be introduced as
\begin{equation}
 \Hil_{\top} = \Oplus_{n=0}^{\infty} \Hil_{\top^n}.
\label{Fock}
\end{equation}

\subsection{QLGA and QCA}

For antisymmetric product \Eq{Fock} has only finite number of terms.
For lattice with $l$ nodes and $m$-dimensional internal space, there are $lm+1$
terms and $\dim \Hil_{\wedge} = 2^{l m}$. So there is some difference 
with cellular automata with same lattice $l$ and internal space $\Hil_S$,
because dimension of Hilbert space of such CA is $\dim \Hil_L = m^l$.
Simplest identification is possible for a case $m=1$ for lattice gas
and $m'=2$ for cellular automaton ($\dim = 2^l$)
\begin{equation}
 \Hil_{\wedge} \cong \Hil_L,
\qquad \Hil_{\wedge} = \Oplus_{n=0}^{l} \Hil_{\wedge^n},
\quad \Hil_L = \Ox_{i=1}^l \Hil_i.
\end{equation}
Similarly, an ``antisymmetric'' lattice gas with arbitrary $m$ may be formally 
represented either by cellular automaton with $m'=2$, but with lattice extended 
by one new dimension $L' = L \times m$, or CA with same lattice $L$ and $m'=2^m$.

For symmetric case it is also possible to use similar transition from lattice
gases to cellular automata. For example instead of lattice gas with $m=1$ and
lattice $L$, it is possible to consider CA with same lattice, but $m = \infty$,
here state of node in lattice is some $N \ge 0$, representing number of particles
in given state. 

Formally such transition from lattice gases to cellular automata in quantum case
is equivalent to construction of ``the Fock space for each cell'' 
\begin{equation}
\Hil_{F(S)} = \Oplus_{n=0}^{\infty} \Hil_{S^n}
\end{equation}
instead of \Eq{Fock} with Fock space for whole lattice and it may be
disadvantage of such CA picture, if to recall global character of
Fock space. It is especially clear, if to try to make some calculations
in ``discrete momentum space'' related with initial lattice coordinates
by discrete Fourier transform.
 
\subsection{Space-Time Model of QLGA}
\label{Sec:qlgt}

For addition of {\em the time dimension} it is necessary instead of spatial
lattice $\ell$ to consider space-time lattice $\ell' = \ell \times \Evol$, {\em i.e.,}
to extend space $\Hil_\ell$ of each qubot, 
$\Hil_{\ell'} = \Hil_\ell \otimes \Hil_\Evol$.

It is possible to write
\begin{equation}
 \Hil_\bot = \Hil_{\ell} \otimes \Hil_S \cong \C^{n_\ell n_S},
\quad
\Hil_\bot^\Evol = \Hil_{\ell'} \otimes \Hil_S \cong \C^{n_\ell n_t n_S},
\end{equation}
where $n_\ell = \dim \Hil_{\ell}$ is number of points in initial lattice, {\em e.g.,}
$n_\ell = n_x n_y n_z$, $n_t$ is number of points (steps) in time dimension,
and $n_S = \dim \Hil_S$ --- dimension of internal space of qubot.

So for one-qubot state some analogue of \Eq{plusT} is really hold, because
\begin{equation}
n_t \to n_t+1 \colon \quad
 \Hil_\bot^\Evol \to \Hil_\bot^{\Evol+1} \cong \Hil_\bot^\Evol \oplus \Hil_\bot.
\end{equation}

On the other hand, let us consider Fock space with all possible antisymmetric
products with different number of qubots
\begin{equation}
 \Hil_{\wedge}^\Evol = 
 \!\!\bigoplus_{k=0}^{n_{\ell} n_t n_S}\!\!\bigl(\Wedge_{i=1}^k \Hil_\bot^\Evol\bigr),
 \quad
 \dim \Hil_{\wedge}^\Evol = 2^{n_\ell n_t n_S}.
\label{aequbot}
\end{equation}
The space \Eq{aequbot} may be formally identified with QCA with same
space-time lattice $\ell'$ and with number of states $2^{n_S}$ for each
site of lattice. So here is held an analogue of \Eq{tensT}, because
simple calculation of dimension shows
\begin{equation}
n_t \to n_t+1 \colon \quad  
\Hil_\wedge^\Evol \to \Hil_\wedge^\Evol \otimes \Hil_\wedge.
\end{equation}
On the other hand, finding of direct equation for evolution of such
a QCA starting with initial QLGA and \Eq{aequbot} looks rather nontrivial.

\subsection{Simplest Example of QLGA to QCA Conversion}
\label{Sec:QLG2CA}

\noindent
\parbox[b]{0.22\textwidth}{
\includegraphics[scale=0.75]{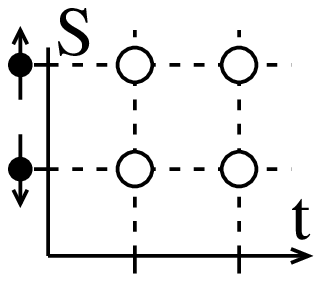}
}~
\parbox[b]{0.75\textwidth}{
Let us consider qubot with two states on lattice with one site and only
two time steps. Let us use a simple scheme for notation depicted by
presented diagram of four-dimensional Hilbert space 
$\Hil_\bot = \Hil_{\ell'} \otimes \Hil_S$. 
Basic vectors of the one-qubot space are denoted as
$\ket{\OI\OO}, \ket{\IO\OO}, \ket{\OO\OI}, \ket{\OO\IO}.$
\medskip
}

For example state 
$
(\ket{\IO\OO}+\ket{\OO\OI})/\sqrt{2}
$ 
corresponds to one qubot evolution
with state $\ket{{\uparrow}}$ at $t=0$ and $\ket{{\downarrow}}$ at $t=1$.
It is ``additive'' scheme \Eq{PlusLatH}.
Two-steps evolution without change of state may be described as
\begin{equation}
 \Ket{\psi_\bot} =
 \alpha\,\bigl(\ket{\OI\OO}+\ket{\OO\OI}\bigr)+
 \beta\,\bigl(\Ket{\IO\OO}+\Ket{\OO\IO}\bigr)
\label{TrivEv2}
\end{equation}

The antisymmetric Fock space may be decomposed 
\begin{equation}
\Hil_\top = \C \oplus \Hil_\bot \oplus \Hil_\bot {\wedge} \Hil_\bot \oplus
\Hil_\bot {\wedge} \Hil_\bot {\wedge} \Hil_\bot \oplus 
\Hil_\bot {\wedge} \Hil_\bot {\wedge} \Hil_\bot {\wedge} \Hil_\bot.
\label{BotsFock}
\end{equation}

To identify the space of QLGA with QCA, let us use the ``spacetime'' lattice with
two sites for moments $t=0$ and $t=1$. In ``multiplicative'' scheme \Eq{TensLatH} 
Hilbert space $\Hil_\bullet$ of each site has four states
\begin{equation}
\ket{\EMP} = \ket{\OO},
\ \ket{\UP} = \ket{\IO},
\ \ket{\DN} = \ket{\OI},
\ \ket{\UD} = \ket{\II}.
\label{SiteFock}
\end{equation}
The states describe ``Fock space of site''. Here ``spatial'' lattice has only one site
and so $\Hil_L = \Hil_\bullet$.
Now Hilbert space $\Hil_\top$ of Fock space \Eq{BotsFock} for system with varying
number of qubots may be described as
\begin{equation}
 \Hil_\top \cong \Hil_L^\Evol = \Hil_L \otimes \Hil_L.
\label{Evol2}
\end{equation}

Due to \Eq{Evol2} each element of \Eq{BotsFock} may be described as tensor
product of two states \Eq{SiteFock}, {\em e.g.,} 
$\ket{\IO\OO}=\ket{\IO}\ket{\OO}= \ket{\UP,\EMP}$, 
$\ket{\IO\OI}=\ket{\IO}\ket{\OI}= \ket{\UP,\DN}$, {\em etc.}

So trivial QLGA evolution \Eq{TrivEv2} may be rewritten for QCA as
\begin{equation}
 \ket{\psi_\bot} =
 \alpha\,\bigl(\ket{\DN,\EMP}+\ket{\EMP,\DN}\bigr)+
 \beta\,\bigl(\ket{\UP,\EMP}+\Ket{\EMP,\UP}\bigr)
\label{TrivEv2CA}
\end{equation}

On the other hand, basic vectors in each term $\Lambda^k_4 \Hil_\bot$ of \Eq{BotsFock} 
may be depicted as
\begin{eqnarray*}
\C \cong \Lambda^0_4 \Hil_\bot & (\dim=1)~\colon & \Ket{\OO\OO} \\
\Hil_\bot \cong \Lambda^1_4 \Hil_\bot &(\dim=4)~ \colon &
\Ket{\OI\OO},~\Ket{\IO\OO},~\Ket{\OO\OI},~\Ket{\OO\IO} \\
\Lambda^2_4 \Hil_\bot &(\dim=6)~ \colon &
\Ket{\OI\OI},~\Ket{\IO\IO},~\Ket{\IO\OI},~\Ket{\OI\IO},
~\Ket{\II\OO},~\Ket{\OO\II}\\
\Hil_\bot \cong \Lambda^3_4 \Hil_\bot &(\dim=4)~ \colon &
\Ket{\II\IO},~\Ket{\II\OI},~\Ket{\IO\II},~\Ket{\OI\II}\\
\C \cong \Lambda^4_4 \Hil_\bot &(\dim=1)~ \colon &
\Ket{\II\II}
\end{eqnarray*}

Antisymmetric (Grassmann) product of basic elements from $\Lambda^k_4\Hil_\bot$ may
be calculated using associativity of the operation $\wedge$ and ``union'' rule
\begin{equation}
\newcommand\m{\phantom{-}}
\renewcommand{\arraystretch}{1.25}
\begin{array}{|c||c|c|c|c|}\hline
\bigwedge   &\m\Ket{\OI\OO}&\m\Ket{\IO\OO}&\m\Ket{\OO\OI}&\m\Ket{\OO\IO}\\ \hline\hline
\Ket{\OI\OO}& 0           &\m\Ket{\II\OO}&\m\Ket{\OI\OI}&\m\Ket{\OI\IO}\\ \hline
\Ket{\IO\OO}&-\Ket{\II\OO}& 0           &\m\Ket{\IO\OI}&\m\Ket{\IO\IO}\\ \hline
\Ket{\OO\OI}&-\Ket{\OI\OI}&-\Ket{\IO\OI}& 0           &\m\Ket{\OO\II}\\ \hline
\Ket{\OO\IO}&-\Ket{\OI\IO}&-\Ket{\IO\IO}&-\Ket{\OO\II}& 0           \\ \hline
\end{array}\qquad
\label{LMulTab}
\end{equation}

\begin{equation}
\newcommand\m{\phantom{-}}
\renewcommand{\arraystretch}{1.25}
\begin{array}{|c||c|c|c|c|c|c|}\hline
\bigwedge   &\m\Ket{\II\OO}&\m\Ket{\OI\OI}&\m\Ket{\OI\IO}
            &\m\Ket{\IO\OI}&\m\Ket{\IO\IO}&\m\Ket{\OO\II}\\ \hline\hline
\Ket{\OI\OO}& 0            & 0            & 0             
            &\m\Ket{\II\OI}&\m\Ket{\II\IO}&\m\Ket{\OI\II}\\ \hline
\Ket{\IO\OO}& 0            & -\Ket{\II\OI}& -\Ket{\II\IO}           
            & 0            & 0            &\m\Ket{\IO\II}\\ \hline
\Ket{\OO\OI}&\m\Ket{\II\OI}& 0            & -\Ket{\OI\II}
            & 0            & -\Ket{\IO\II}& 0 \\ \hline
\Ket{\OO\IO}&\m\Ket{\II\IO}&\m\Ket{\OI\II}& 0
            &\m\Ket{\IO\II}& 0            & 0 \\ \hline
\end{array}
 \tag{\ref{LMulTab}$'$}
\label{LMulTab2}
\end{equation}
\begin{equation}
\Ket{\II\II} = \Ket{\OI\OO}\Wedge\Ket{\IO\II}= -\Ket{\IO\OO}\Wedge\Ket{\OI\II}
=\Ket{\OO\OI}\Wedge\Ket{\II\IO} =-\Ket{\OO\IO}\Wedge\Ket{\II\OI}
\tag{\ref{LMulTab}$''$}
\label{LMulTab3}
\end{equation}

Now it is possible to calculate trivial evolution of two qubots, each one is 
described by \Eq{TrivEv2} with different pair of coefficients
\begin{equation}
 \ket{\psi_{\bot\mathnormal 1}} \wedge \ket{\psi_{\bot\mathnormal 2}} =
 (\alpha_1 \beta_2 - \alpha_2 \beta_1)\,
\bigl(\Ket{\II\OO}+\Ket{\OO\II}-\Ket{\IO\OI}+\Ket{\OI\IO} \bigr).
\end{equation}
So for equal states it is zero and for nonequal qubots after normalization 
it is always the same ``Bell-like'' state
\begin{equation}
\frac{1}{2}\bigl(\Ket{\UD,\EMP}+\Ket{\EMP,\UD}+\Ket{\DN,\UP}-\Ket{\UP,\DN}\bigl).
\label{TrivTwo}
\end{equation}


\end{document}